\documentclass[aps,prd,twocolumn,letterpaper,floatfix,showpacs,amsmath]{revtex4} 
\newlength{\picwidth}

\usepackage[dvips]{graphicx}

\newcommand{\bj}[1]{\mbox{\boldmath $#1$}} 
\newcommand{\uj}[1]{\mbox{ #1}}

\newcommand{\inStar}[0]{\Sigma \uparrow 0}
\newcommand{\outStar}[0]{\Sigma \downarrow 0} 
\newcommand{\vvec}[0]{\mbox{\boldmath$v$}}
\newcommand{\di}[0]{\mbox{\boldmath$\nabla\cdot$}}
\newcommand{\gr}[0]{\mbox{\boldmath$\nabla$}}

\begin{document}

\renewcommand{\thefigure}{\arabic{figure}}
\title{
A Nonlinear Coupling Network to Simulate the Development of the r-mode Instablility in Neutron Stars
I. Construction
}
\author{Jeandrew Brink}
\affiliation{Center for Radiophysics and Space Research,
Cornell University, Ithaca NY 14853}
\author{ Saul A. Teukolsky}
\affiliation{Center for Radiophysics and Space Research,
Cornell University, Ithaca NY 14853}
\author{Ira Wasserman }
\affiliation{Center for Radiophysics and Space Research,
Cornell University, Ithaca NY 14853}

\begin{abstract}

R-modes of a rotating neutron star are unstable because of the emission of gravitational radiation. We explore the saturation amplitudes of these modes determined by nonlinear mode-mode coupling. Modelling the star as incompressible allows the analytic computation of the coupling coefficients. All couplings up to $n=30$ are obtained, and  analytic values for the shear damping and mode normalization are  presented. In a subsequent paper we perform numerical simulations of a large set of coupled modes.

\end{abstract}

%\keywords{GR}%Use showkeys class option if keyword
                              %display desired
\maketitle

\section{Introduction}
\label{sec:intro}
In 1998,  Andersson \cite{Anderson}, and Friedman and Morsink \cite{FriedmanMor} showed that r-modes of a rotating neutron star can be unstable via the gravitational-radiation driven Chandrasekhar-Friedman-Schutz (CFS) mechanism \cite{Chandra1970} \cite{FriedmanShutz1978}. In inviscid stars this mechanism  causes any mode that is retrograde in the co-rotating frame but prograde in the inertial frame to grow as it emits gravitational radiation.

For neutron star rotational frequencies $\sim 100-1000$Hz  gravitational radiation from an unstable r-mode falls within the range in which ground-based gravitational wave detectors such as LIGO are sensitive. If detectable this signal  would  probe  the internal structure of neutron stars. Moreover, it has been suggested that the CFS instability of the r-mode could limit neutron star rotational frequencies \cite{Chakrabarty}. The key question, which currently remains unanswered, is to what amplitude these modes grow and whether they can be detected at all.

The slow growth rate of these secularly unstable modes, as well as their three-dimensional nature, makes the problem ill-suited to direct hydrodynamical simulation.  Gravitational radiation slowly drains away the rotational energy of the star over a time scale well in excess of $10^6$ rotation periods, too long a time period for numerically stable hydrodynamical simulations of a rotating neutron star.

There have been three attempts at direct hydrodynamical calculations \cite{Stergioulas} \cite{LindblomEvolve} \cite{Ster2}. These simulations (1) assume that the background star has uniform initial rotation, (2)  use approximate linear eigenfunctions as initial perturbations, and (3)  assume either a large initial amplitude, or else artificially increase the radiation reaction force enough to ensure that a large amplitude is reached relatively quickly.

In one direct hydrodynamical calculation, Stergioulas and Font \cite{Stergioulas}  used $116^3$ grid points, and perturbed an equilibrium model with an excitation corresponding to an approximate linear $l=m=2$ r-mode eigenfunction with a substantial amplitude ($\alpha = 1$).
They evolved the system in a fixed, static spacetime geometry for 26 r-mode periods, after which the amplitude decreased because of numerical viscosity. No energy leakage out of the r-mode was observed, so they concluded  that nonlinear couplings would not inhibit the instability.

Lindblom et al \cite{LindblomEvolve} did a Newtonian hydrodynamical simulation on a $64\times 128\times 128$ grid, with the gravitational radiation  added as a radiation reaction force. However, 
they boosted the radiation reaction force by a factor of 4500 to force the instability to grow substantially in the $\sim 30$ rotation periods spanned by their calculation.  The initial condition was a uniformly rotating star with a small amplitude perturbation proportional to the vector spherical harmonic $Re (Y^B_{22})$.  
In their simulation, the mode amplitude grew exponentially at first, reaching an amplitude of order unity. After that, the growth was halted by dissipation resulting from breaking waves (shocks).

Gressman et al \cite{Ster2} performed a Newtonian hydrodynamical simulation for non-viscous fluid flow very similar to \cite{LindblomEvolve} using  a high resolution shock capturing scheme to accurately integrate their equations. They began by amplifying an initially weak  ($l=m=2$) r-mode at $9\times 10^7$ times the radiation reaction rate, to amplitudes of order unity, at which point they ceased driving the mode and observed its decay. Subsequently the  r-mode decreased in amplitude slowly for a short time,  but then decayed catastrophically. Gressman et al interpreted the dropoff as a slow leakage of energy to other fluid modes,  which then act nonlinearly with the r-mode, resulting in the rapid decay. The time to decay was found to be amplitude-dependent: the larger the initial amplitude the faster the eventual decay. Previously, we have argued that this behavior can be attributed to the non-linear period of a particular three-mode coupling \cite{JdB3}.

None of these simulations excludes the possibility that the r-mode saturates at low amplitudes. If the timescale for the growth of the ($l=m=2$) r-mode to saturate by nonlinear hydrodynamical effects is longer than the unphysically shortened growth  rates used in the simulations,  saturation at small amplitudes could  be prevented artificially. Moreover, the spatial resolution of the hydrodynamic codes may not be sufficient to resolve the small scale modes to which the r-mode couples effectively, artificially increasing the possible r-mode amplitude.

To gain a deeper understanding of the r-mode instability and its saturation amplitude, 
we investigate the nonlinear coupling of a large network of the inertial modes of a  star.  As long as the amplitudes of the modes remain small, it is a good approximation to treat the oscillations with (weakly nonlinear) perturbation theory.  By treating the neutron star as a uniform density incompressible object, the modes and their couplings can be calculated analytically.
Our dynamical model then reduces to the simultaneous solution of a network of equations of the form \cite{Katrin1}
\begin{equation}\dot{c}_A(t) - iw_Ac_A +\gamma_Ac_A = -i\frac{w_A}{\epsilon_A} \sum_{BC}\kappa_{\overline{A}BC} c_B c_C
\label{eq:ampeq}
\end{equation}
Here the amplitudes $|c_A|$ are assumed to be small, so we retain only the nonlinear couplings among triplets of modes, which, in the absence of dissipation and driving, would form an infinite dimensional Hamiltonian system with three-mode couplings $\kappa_{\overline{A}BC}$. Gravitational radiation and viscosity are added as driving and damping terms $\gamma_A$.  The quantities $w_A$ and $\epsilon_A$ are the mode frequency and mode energy of mode A respectively and a bar denotes complex conjugation. 

This approach has the great advantage that we can track the energy transfer % mechanism 
 and identify the couplings that become important. 
However it also poses many analytical and computational challenges.  We present the analytic foundation to this investigation in this paper. The numerical investigation of the evolution of the oscillator network and the saturation amplitudes achieved will be presented in a later paper \cite{JdB2}.  

First the linear eigenfunctions have to be obtained  and appropriately normalized.  In the case of a uniformly rotating uniform-density star an elegant coordinate transformation introduced by Bryan yields an exact analytic solution \cite{Bryan} \cite{Ipser}. A few properties of these generalized r-modes, their frequency structure and analytic normalization are reviewed in Section \ref{sec:Bryan}.

 Section \ref{sec:DampDrive}   details damping and driving  terms of the linear modes. The analysis is a brief summary of that presented by Lockitch and Friedman\cite{DampLock} and  Lindblom, Owen and Morsink \cite{DampLind} and serves as a partial test of our approach to computing integrals over the star. 
%Only  damping due to shear viscosity is considered and 
Analytic scalings for the damping are obtained in the zero rotation limit. 

The main body of this paper explores a method of computing the nonlinear mode-mode couplings.  In Section \ref{CouplingCoeff} we specialize the formalism presented by Schenk et al \cite{Katrin1} to the case of the inertial modes of an incompressible star. Computing the coupling coefficients for modes with large mode number $n=l+1$ is tricky because of precision problems. A  technique for computing these coefficients  to arbitrary precision is presented in Section \ref{CompMethod}. Computational time however limits us to computing only coefficients up to $n = 30$.  Section \ref{Selectionrules} explores some of the properties of the coupling coefficients, including a complete list of selection rules they obey.

Arras et al \cite{Saturation}  based their WKB analysis on the Bryan modes to determine the scalings of  
the damping, normalizations and couplings in their weak turbulence picture of r-mode saturation. Whereas they employed the Cowling approximation in  computating the coupling coefficients, we do not.  
In fact, the perturbations of the gravitational potential make a nonnegligible contribution to the coupling coefficients in this problem.

Arras et al \cite{Saturation} also presented a picture where a cascade of energy from the r-mode to smaller length scales is set up, and an inertial range is established.  Our papers are designed to test this picture ``experimentally'',  and to see whether an alternative scenario is valid where the r-mode is effectively damped by coupling to just a few modes.

\section{Equilibrium solution and Linear Modes}
\label{sec:Bryan}

Consider a rotating ball of liquid with constant density that obeys the Euler equation 
\begin{equation}\partial_t \vvec +\vvec \cdot \gr \vvec = -\gr \left(\frac{p}{\rho}-\Phi\right)
\end{equation} the continuity equation 
\begin{equation}\partial_t \rho + \di (\rho \vvec)=0
\end{equation}
and Newtonian gravity 
\begin{equation}
\nabla^2 \Phi = -4\pi G\rho \end{equation} 
Here, \vvec ~ is the velocity field, $\rho$ the density, $\Phi$ the gravitational potential of the star and $p$ the pressure.

 Maclaurin spheriods are 
uniform density,  uniformly rotating solutions with  eccentricity $e$ and volume $V=\frac{4}{3}\pi R^3$. The average radius of the spheroid $R$ is related to the equatorial ($R_e$) and  polar ($R_p$) radii of the spheroid by $R^3=R_e^2R_p$.  The pressure and angular velocity of the spheroid can be expressed in terms of the eccentricity as follows:

\begin{align}
p&= p_e \left(1 - \frac{\varpi^2}{R_e^2} -
\frac{z^2}{R_p^2}\right) \label{eq:p}\\
\Omega^2 &= 2\pi G \rho\left(\frac{(3-2e^2)\sqrt{1-e^2}}{e^3}\arcsin e - \frac{3}{e^2}(1-e^2)\right)\label{eq:OME}
\end{align}
where $G$ is Newton's Constant and $\varpi^2 = x^2+y^2$ is the cylindrical radial coordinate. 
 For a given eccentricity, $p_e$, $R_e$ and $R_p$ are constant:
 \begin{eqnarray}
p_e& =& 2\pi G\rho^2  \frac{ (1 - e^2)^{2/3}}{e^2}\left(1
-\sqrt{\frac{1 - e^2}{e^2}}\arcsin e \right)\\
R_e &=&R\left(\frac{1}{1-e^2}\right)^{1/6}\\
R_p &=&R\left(1-e^2\right)^{1/3}~.
\end{eqnarray}

These solutions are dynamically and secularly stable below the bifurcation point  at  $e = 0.81267$, $\Omega = 0.61174 \sqrt{\pi G\rho}$ \cite{Ellips}.  

\subsection{Linear Eigenmodes}

The linear stability analysis was carried out by Bryan in 1889 \cite{Bryan}. 
Ipser and Lindblom \cite{Ipser} have given a  modern and improved rendering of this analysis and it is mainly their approach and notation on which we rely.

The linearized perturbation equations for the Lagrangian displacement \bj{\xi} are 
\begin{eqnarray}
\bj{\ddot{\xi}}+2\Omega\times \bj{\dot{\xi}} = -\nabla \delta U \label{eq:xievolve}\\
\nabla^2 \delta \Phi = - 4\pi G \delta \rho ~. \label{eq:grav}
\end{eqnarray}
The Coriolis force is represented by the $2\Omega\times \bj{\dot{\xi}}$ term. $\delta U$ is the hydrodynamic potential of the star,  related to the Eulerian pressure perturbation $\delta p$ and gravitational potential perturbation $\delta \Phi$ by 
\begin{equation}
\delta U = \frac{\delta p}{\rho}-\delta \Phi~. 
\label{Eulerian_p}
\end{equation}
 Assuming an $e^{i\omega t}$ time dependence, (\ref{eq:xievolve}) and (\ref{eq:grav}) can be rewritten using the two-potential formalism of   Ipser and Lindblom \cite{Ipser} (this frequency convention differs from that used by Schenk et al \cite{Katrin1} by a minus sign):
\begin{eqnarray}
\nabla^2\delta U& =& \frac{1}{w^2}\partial_{zz}\delta U \label{eq:dU}\\
\nabla^2 \delta \Phi &=& -4\pi G \rho^2 \delta(p) \left[ \delta U+\delta \Phi\right]~. \label{eq:dPhi}
\end{eqnarray}

Note that for the rest of this paper the rotational frequency $\omega$ will be rescaled in terms of the parameter $w =\omega/2\Omega$ which takes on values between -1 and 1.  ($w$ is related to  the $\kappa$ frequency parameter of Ipser and Lindblom \cite{Ipser} by $w=\kappa/2$.)

Equation (\ref{eq:dPhi}) for the gravitational potential is Laplace's equation with a discontinuity on the boundary. The delta function describes the surface of the Maclaurin spheroid. Equation \eqref{eq:dU} for $\delta U$ is valid only within the spheroid.

With two elegant coordinate transformations, Bryan transformed equations (\ref{eq:dU}) and (\ref{eq:dPhi}) so that they become separable and allow the boundary conditions to be imposed. On the surface of the star these eigenfunctions correspond to waves that preserve the stellar volume. Imposing the boundary conditions leads to dispersion relations for the eigenfrequencies.

The eigenfunctions found by Bryan \cite{Bryan},\cite{Ipser} can be expressed as Legendre functions in spheroidal
coordinates $(\zeta,\mu)$ for the gravitational potential
\begin{eqnarray}
\delta \Phi_{\outStar}& =& \alpha
\frac{Q^m_n(i\zeta)}{Q^m_n(i\zeta_0)}P^m_n(\mu)e^{im\phi} \nonumber\\
\delta \Phi_{\inStar}& =& \alpha
\frac{P^m_n(i\zeta)}{P^m_n(i\zeta_0)}P^m_n(\mu)e^{im\phi}\ . \label{dPhiEig} 
\end{eqnarray}
Here $\alpha$ is an arbitary constant 
and  $\zeta_0 = (1-e^2)/e^2$. The subscripts $_{\outStar}$  and $_{\inStar}$ indicate the exterior and interior regions of the star respectively.  
The coordinate transformation to spheroidal coordinates is 
\begin{align}
&\varpi = a\sqrt{(\zeta^2 + 1)(1-\mu^2)}& z &= a\zeta\mu
\end{align}
where the constant $a$ is related to the eccentricity by $a^3=R^3e^3/\sqrt{1-e^2}$

The hydrodynamic potential equation (\ref{eq:dU}) is solved
in terms of bispheroidal coordinates~$(\xi,\tilde{\mu})$:
\begin{align}
\varpi &=
b\sqrt{(1-\xi^2)(1-\tilde{\mu}^2)}& z &=bd\xi\tilde{\mu} \notag \\
b^2&=\frac{a^2(1-w^2e^2  )}{(1-w^2)e^2}&d&=\frac{\sqrt{1-w^2}}{w} \label{eq:BiSpheroidal}
\end{align}
and
\begin{eqnarray}
\delta U &=& \beta\frac{P^m_n(\xi)}{P^m_n(\xi_0)}P^m_n(\tilde{\mu})e^{im\phi} 
\label{dUEig}
\end{eqnarray}
where  $\beta$ is a constant related to $\alpha$  by the dispersion relations due to the boundary conditions. The constant $\xi_0^2 = a^2\zeta^2_0/b^2d^2 $ is the coordinate value of the stellar surface  in bispheroidal coordinates. Note that the Legendre functions are labeled using the indices $n$ and $m$.  This is to avoid confusion with the index $l=n-1$ that is used to denote the spherical harmonics in the radiation reaction force, equation \eqref{GRDrive}.

The  Lagrangian displacement $\bj{\xi}$ can be constructed from the hydrodynamic potential \cite{Ipser} equation (3.2).
\begin{equation} 
\bj{\xi}^a = \frac{1}{4 \Omega ^2} \overline{Q}^{ab} \delta U_{;b}
\label{eq:BryanEig}
\end{equation}
where 
\begin{equation}
 \overline{Q}^{ab} = \frac{1}{w^2-1}(g^{ab}-\frac{1}{w^2}z^a z^b - \frac{i}{w}\nabla^a\phi^b) 
\end{equation}

Finally the Lagrangian perturbation in pressure, equation (I5) of  \cite{Katrin1} \mbox{$\Delta p = \delta p +\bj{\xi}\cdot \nabla p$}  can be written as 
\begin{eqnarray}
\Delta p& =&\rho(\delta U+\delta \Phi) +\bj{\xi}\cdot \nabla p \label{Lagrangep}
\end{eqnarray} 
using equation  \eqref{Eulerian_p}.
The boundary conditions require that  $\Delta p = 0$ on the surface of the star. For the special case of the $n=m+1$ r-modes it can be shown (see equation (7.11) of \cite{Ipser}) that  $\Delta p$ is identically zero over the entire star for any rotation rate.

\subsection{Properties of Linear Modes}
The coordinate transformations \eqref{eq:BiSpheroidal} used   for the hydrodynamic potentials depend on the mode frequency, $w$.  Later, when the coupling coefficients are computed in Section \ref{CouplingCoeff}, it will be necessary to perform integrals involving three different modes; this requires a common coordinate system. It turns out that when transformed back to cylindrical  coordinates, the modes are nothing more than polynomials in $\varpi$ and $z$ 
with frequency dependent coefficients, which makes the problem tractable.
The construction of these polynomials is achieved by successive applications of recursion relations of the Legendre functions; this is presented in Appendix~\ref{BryanRecursion}.  The fact that the potentials can be represented in this way allows us to compute the large number of coupling coefficients between the modes in a very efficient and accurate manner.

Each eigenfunction can be labeled by three numbers $(n,m,k)$ varying over the ranges specified below:
\begin{eqnarray}
n:& 2\rightarrow \infty &\mbox{Principal Legendre index} \nonumber \\
m:& 0\rightarrow n-1 & \mbox{Azimuthal index} \nonumber \\
k:& 1\rightarrow n-m\ \mbox{if}\ m\neq 0 & \mbox{Frequency index.} \nonumber\\
 & 1\rightarrow n-m-1\ \mbox{if}\ m = 0 & \nonumber 
\end{eqnarray}

In order to estimate the number of modes for a given $n$, and the number of couplings among these modes, it is useful to introduce the index $j$ which arranges the inertial modes consecutively: 
\begin{equation} j = \frac{(n-1)n(n+1)}{6} +  \frac{(n-m-1)(n-m)}{2} +  k-1~.\end{equation}
The $j$ indices for the $n=m+1$ r-modes are \mbox{$j_R =  (n-1)n(n+1)/6$}.

To give some idea of the size and connectivity of our coupling network, observe that for a given maximum value of $n=n_{max}$ , the number of included modes is 
\begin{equation}
N_{max} = \frac{n_{max}(n_{max}+1) (n_{max}+2) }{6} -1 \approx \frac{n_{max}^3}{6}.
\end{equation}

 The selection rules discussed in Section~\ref{Selectionrules} restrict the total number of couplings ($K$) to 
\begin{equation} K \propto n_{max}^8.\end{equation}

\subsection{Frequency Spectrum}
\begin{figure}
\includegraphics[width=\columnwidth]{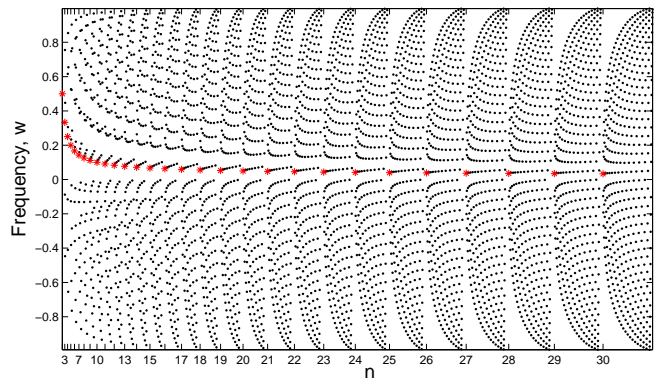}
\caption{ Frequency spectrum for the inertial modes of an incompressible star. $n=m+1$ r-modes denoted by `*'. The $x$-axis ticks denotes the mode number $n$. The data points for the various $m$ values follow each tick starting with the mode $m = n-1$ nearest the tick mark and continuing with $n-2 \dots 0$. The $y$-axis plots the frequency value $w= \omega/2\Omega$. } 
\label{Freqfig}
\end{figure}

The frequency spectrum for the Bryan modes is set by the boundary conditions \cite{Ipser} and  is shown in Figure~\ref{Freqfig}. The physical values of these frequencies are bounded between $-2\Omega$ and $2\Omega$. (Numerical values for the first few frequencies can be found in \cite{Ipser} Table I )

For comparison, the oscillation frequency  ($f$-mode) of an  incompressible fluid sphere (\cite{Bryan} equation (86)) is \begin{equation}\omega_f^2 = \pi G \rho \frac{8n(n-1)}{3(2n+1)}~.\end{equation}
More generally,  the $f$ and   $p$ modes of a compressible star have frequencies scaling as $\sqrt{\pi G \rho}$.  These frequencies are much larger than the r-mode frequencies even for moderately fast rotation.
The large separation between these two frequency regimes makes the evolution of the r-mode difficult to follow in a hydrodynamical calculation.  Any direct fluid simulation requires a time step small enough to resolve the oscillations due to rapid pulsation modes, while studying the r-mode problem requires the evolution of  modes with much longer timescales.  The large discrepancy between the frequencies of inertial and pulsation modes makes it unlikely that the generalized r-modes would excite the pulsation modes. In all of our subsequent computations the pulsation modes will be ignored, allowing much larger time steps to be taken in mode-mode simulations, and enabling us to study the long term effects of mode-mode interactions.

\subsection{Normalization}
To facilitate the comparison between modes, all the linear modes are  normalized by a factor $\psi$ such that  the mode rotating-frame-energy at unit amplitude, $\epsilon$, has a fixed value of $ 2 E_{unit} =MR^2\Omega^2$ where $M=\frac{4}{3}\pi\rho R^3$ is the mass of the star. 
  
Schenk et al. \cite{Katrin1} equations (K22) and (2.36) provide a general formula for the mode energy, valid at any rotation rate, namely 
\begin{eqnarray}
\epsilon&=&-\omega \psi^2 \left(i\left<\bj{\xi},\bj{2\Omega}\times\bj{\xi}\right>-2\omega\left<\bj{\xi},\bj{\xi}\right>\right)\nonumber\\
&=&\psi^2w^2\int d^3x\ \rho \bj{\xi^*}\cdot\left(\frac{\nabla \delta U}{w^2} +4\Omega^2\bj{\xi}\right)\label{eq:ModeEnergy}
\end{eqnarray}
These integrals can be computed analytically for an incompressible star of any ellipticity.  The computation  involves a number of nonstandard integrals of Legendre functions \cite{JdBThesis}.
In the spherical limit in units where $R = 1$ the  mode energy greatly simplifies to
\begin{eqnarray}
\epsilon&=&\frac{\psi^2}{4\Omega^2} \frac{2 \pi \rho \beta^2 }{(1-w^2)}  \frac{2}{2n+1}\frac{(n+m)!}{(n-m)! }\left[   n(n+1)\right]~. 
\end{eqnarray}
and the dimensionless normalization factor $\psi$ is 
\begin{eqnarray}
\psi &=& 4\Omega^2\frac{1}{\beta}\sqrt{\frac{\pi}{3}}
 \sqrt{\frac{1-w^2}{n(n+1)}}\sqrt{\frac{2n+1}{4\pi}\frac{(n-m)!}{(n+m)!}}~.
\end{eqnarray}
The last square root factor is the standard normalization factor of Legendre functions to obtain spherical harmonics.

\section{Dissipative Effects}
\label{sec:DampDrive}
If the dissipative effects are small they can be modeled by adding a damping term $\gamma_Ac_A$ to the oscillator equation (\ref{eq:ampeq}). For a detailed discussion of the validity of this approach see \cite{DampLock}.

The main damping mechanism in the higher order modes is viscosity, while gravitational radiation both damps and drives the low-order modes, with very little effect on the higher order modes.

\subsection{Shear viscous damping} 

 The damping rate, $\gamma_\eta$ of a mode due to shear viscosity can be computed from the perturbed shear tensor \cite{DampLock}
\[\delta \sigma^{ab} = iw\Omega\left(\bj{\xi}^{a;b}+\bj{\xi}^{b;a}\right)\]
to be 
\begin{eqnarray}
\gamma_{\eta} = -\frac{\dot{E_{\eta}}}{\epsilon}& = &\frac{2\eta}{\epsilon}\int d^3x\  \delta\sigma^{ab}\delta\sigma_{ab}^*\nonumber\\
&=&  \frac{\eta}{\epsilon}\int d^3x\ (2w\Omega)^2 (\bj{\xi}^{a;b} \bj{\xi}_{a;b}^* +\bj{\xi}^{a;b} \bj{\xi}_{b;a}^*)  \nonumber
\end{eqnarray}
where $\eta$ is the shear viscosity coefficient.  We use the coefficient proposed by \cite{DampLock}\cite{1987ApJ...314..234C}
 for hot neutron star matter, and explore temperature ranges varying from $10^6$K to $10^9$K:

\[\eta = 2\times 10^{18}\left(\frac{\rho}{10^{15} \uj{g cm}^{-3}}\right)^{9/4}\left(\frac{10^9\ \uj{K}}{T}\right)^2 \uj{g cm}^{-1}\ \uj{s}^{-1}\]

The geometric contribution $(\gamma_\eta/\eta)$ of the individual modes is rather complex, having different scalings for a fixed $n$ depending on the $m$ value as well as the frequency. It varies from roughly $2n^2$ for the $m=n-1$ r-modes to $\frac{2}{3}n^3$ for the $m=0$ modes. 
This picture is more complicated than the  scalings given by Arras et al. in \cite{Saturation} based on the WKB analysis

An analytic fit valid for all the modes motivated in part by the partial analytic results for the shear \cite{JdBThesis} 
can be found for our numerically computed shear damping rates:  
\begin{eqnarray}
\frac{\gamma_\eta}{\eta}
&=& \frac{1}{3}(2n+1)\left[(n+3)(n-2) -  \frac{m^2-2mw}{1-w^2} \right]\label{eq:ShearVis}
\end{eqnarray}

\begin{figure*}
\includegraphics[width=\picwidth]{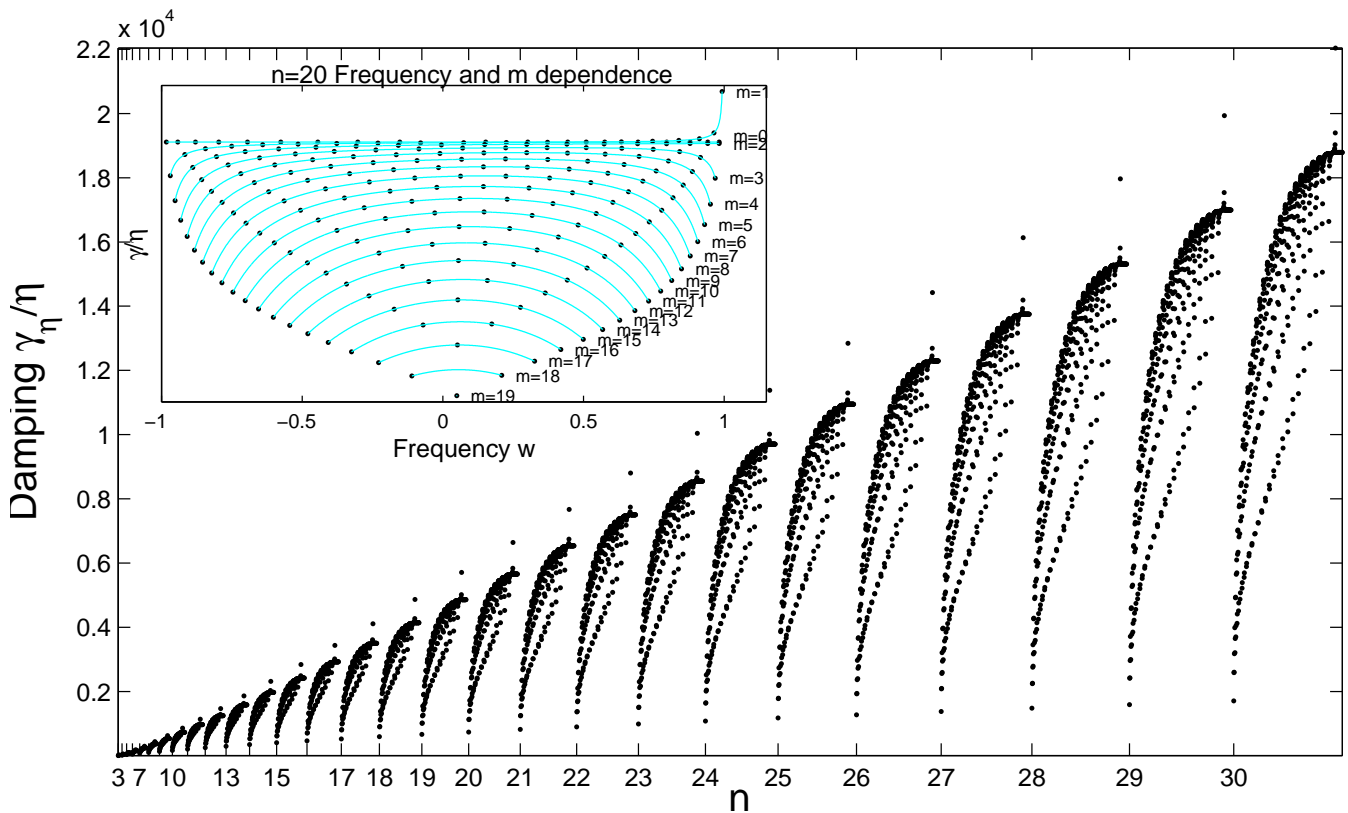}
\caption{ Damping due to shear viscosity of the inertial modes. The $x$-axis ticks denote the mode number $n$. The data points for the various $m$ values follow each tick, starting with the mode $m = n-1$ nearest the tick mark and continuing with $n-2 \dots 0$. The $y$-axis plots the geometric contribution of the shear $\gamma_\eta/\eta$. The inset displays the frequency and $m$ dependence for  $n=20$. Lines join points with a fixed $m$ value and are the calculated using the fit of equation (\ref{eq:ShearVis}).  } 
\label{Dampingfig}
\end{figure*}

For the case of the  $m = n-1$ r-mode the damping reaches its minimum value of
\begin{eqnarray}
\frac{\gamma_\eta}{\eta} 
 &=& (2m+3)(m-1)
\end{eqnarray}
for a given $n$. Figure \ref{Dampingfig}  displays the results for the shear damping.

\subsection{Gravitational driving}
Gravitational radiation strongly drives certain low-order modes, resulting in large scale fluid circulation while damping others. An accurate estimate of the driving rates of gravitational radiation can be found using the procedure outlined by Ipser et al \cite{DampLind} and is summarized in the following equations: 
\begin{equation}
\gamma_{GR}  
= \frac{2w(2w-m)\Omega^2}{2E} \sum_{l\geq 2} N_l (2w-m)^{2l}\Omega^{2l}\left(|\delta D_{lm}|^2+|\delta J_{lm}|^2\right)
\label{GRDrive}
\end{equation}
where $\delta D_{lm}$ is the mass multipole (which vanishes in the spherical limit) \begin{equation}\delta D_{lm} = \int \delta \rho\ r^l\ Y^*_{lm}\ d^3x \end{equation}
and  $\delta J_{lm}$ is the current multipole (which makes the largest contribution to driving in our model),
\begin{equation}\delta J_{lm} = \frac{2}{c} \sqrt{\frac{l}{l+1}} \int  r^l (2w\Omega\rho \bj{\xi} + \delta \rho  \bj{v})\cdot Y^{B*}_{lm} \ d^3x\end{equation}
$Y_{lm}$ and $Y^B_{lm}$ are the scalar and magnetic type vector spherical harmonics respectively.
The quantity 
\begin{equation}
N_l = \frac{4\pi G}{c^{2l+1}}\frac{(l+1)(l+2)}{l(l-1)[(2l+1)!!]^2}
\end{equation} determines the strength of the coupling. The $c^{2l+1}$ factor in the denominator is the reason only large scale, small $n$ modes such as the ($n=3$,$m=2$) r-mode are driven. 

The sign of the damping/driving rate is set by the $2w(2w-m)$ factor. The eigenmode will be driven if this product is less than zero for that particular mode. This is the mathematical statement leading to the CFS instability.

\def\div{{\mbox{\boldmath $\nabla\cdot$}}}
\def\xivec{{\mbox{\boldmath $\xi$}}}

\subsection{Other sources of dissipation}
Bulk viscosity may be significant in neutron stars
at high temperatures, particularly as a consequence
of the appearance of hyperons at high densities \cite{Owen}.
The rate of dissipation of energy due to bulk viscosity \cite{DampLock} can be written using equations~{(I14) and (I17)}  of \cite{Katrin1} as follows

\begin{align}
-\dot E_B=\int d^3x~\zeta\omega^2\left\vert\div
\xivec\right\vert^2 
\left\vert{\Delta\rho\over\rho}\right\vert^2\notag
\\ =\int d^3x~\left({\zeta\omega^2\over\Gamma_1^2}\right)
\left\vert{\Delta p\over p}\right\vert^2
~.
\label{bulkvisc}
\end{align}
Mathematically, this is zero for an incompressible star
because $-\dot E_B\propto\Gamma_1^{-2}\to 0$.
We can adopt the attitude that Eq. (\ref{bulkvisc})
gives the dissipation rate to leading order in $\Gamma_1^{-2}$,
which we then regard as a parameter (which can be absorbed into
the still-uncertain bulk viscosity coefficient $\zeta$). In the limit of slow rotation this integral simplifies to  
\begin{equation}
\begin{split}
-\dot E_B =\left({\zeta\omega^2\over\Gamma_1^2}\right) \int d^3x 
\left\vert{\bj{\xi}\cdot \nabla p\over p}\right\vert^2
\end{split}
\label{bulkviscsmallrot}
\end{equation}
where use has been made of equation~\eqref{Lagrangep} and the fact that $\bj{\xi}$ scales like $(1/\Omega^2) \nabla \delta U$ in equation \eqref{eq:BryanEig}. We have evaluated the integral \eqref{bulkviscsmallrot}  up to modes with
$n=12$ and observe a strong frequency dependence in the results, modes with frequencies near zero being strongly damped. However, a deficiency of the incompressible star
model for inertial mode perturbations is that $\Delta p
=0$ for all  $n=m-1$ r-modes.

Because we are mainly interested in the dynamical aspects
of three mode coupling in saturating the CFS instability
of the r-mode, and since we cannot model bulk viscosity
realistically or uniquely using an incompressible star
model, we have chosen only to include shear viscosity in
our network calculations. Viscosity plays two roles in
our problem: (i) it is important in determining the 
parametric instability thresholds for interacting triplets
of modes; (ii) it limits the range of modes that participate
vigorously in the nonlinearly interacting network. Our goal
during this investigation is to explore the characteristic
behavior of mode-mode coupling in the simplest self-consistent
model we can construct. The importance of bulk viscosity,
as well as any other form of damping, can be examined
retrospectively after we have explored the influence of
damping of different strengths on our illustrative
calculations for networks with shear viscosity alone.
Accordingly, we reserve further comment on the omission of
bulk viscosity to our next paper, where we report on 
the results of detailed numerical investigations.

\subsection{Parameters of the Model}

All physical quantities in the calculation are made nondimensional by expressing mass in units of $M_\odot$, time in units of $1/2\Omega$, and length in units of the mean radius of the star, $R$.

This choice of units allows the coupling coefficients to remain unchanged for all stars to first order in  $\Omega^2$. (Changing the rotation rate of the star influences the eigenfunction and this effect enters the coupling coeffiecients as second order in $\Omega^2$.)  It does however, mean that if the rotation rate of the star is changed both the viscous damping and GR driving rates are changed.  The viscous damping rate changes because time is rescaled. The GR driving rate changes because $\Omega$ enters nonlinearly into the formula for the GR driving.

The dimensionless parameters in our simulations are chosen to correspond to the physical values of a star with $R=10 \uj{km}$, $M=1.4 M_\odot$ and temperature ranging from   $10^6$K to  $10^9$K. The eigenfunctions considered in our calculations are those obtained in the limit of no rotation.  However the gravitational driving is computed as if the star were rotating with a fixed rotation rate of $\Omega = 0.37 \sqrt{\pi \rho G}$ corresponding to an eccentricity of $e=0.5$. This rotation rate is fast corresponding to a physical frequency of $ \nu= 695$Hz or a period of 1.4 ms, which is appropriate since we are attempting to set an upper bound on the possible dynamics due to the r-mode driving.

Our computed values for the damping and driving rates agree with those computed by  Lockitch et al. \cite{DampLock} et al %to 3 decimal places 
for the case of a  neutron star with $R = 12.57 \uj{km}$,  $T = 1\times 10^9 \uj{K}$, $M=1.4 M_\odot$ and  $\Omega^2 = \pi G \rho$, which are the values to which they scaled their results. 

The often-quoted growth time due to gravitational radiation of the order of $10^3$ rotation periods is the {\it maximum} possible growth rate that can be achieved if the star were rotating at breakup. Actual growth times could be orders of magnitude larger and increase like $\Omega ^{-4}$ for the ($n=3$,$m=2$) r-mode .

\section{Nonlinear coupling}
\label{CouplingCoeff}

The linear growth of the unstable modes can be halted by various damping mechanisms. We wish to explore the damping due to couplings to other modes which eventually transfer energy to small enough scales that it is efficiently dissipated.
The pathways available for the energy to dissipate are determined by the couplings and frequency detunings between the modes.
We consider only the lowest order nonlinear three-mode couplings as computed  by 
Schenk et al \cite{Katrin1}. Their explicit expression for three-mode coupling as given by equation (4.20)\cite{Katrin1} is  

\begin{widetext}
\begin{eqnarray}
\kappa_{ABC}& =& \frac{1}{2} \int d^3x\ p (\Gamma_1
-1)  \Xi_{\{AB}\nabla\cdot\bj{\xi}_{C\}}
+  p\left[ (\Gamma_1-1)^2+\frac{\partial\Gamma_1}{\partial \ln
\rho}\right] \nabla\cdot\bj{\xi}_A \nabla\cdot\bj{\xi}_B \nabla\cdot\bj{\xi}_C\nonumber\\
&+& \frac{1}{2} \int d^3x\ p\left(\chi_{ABC}+\chi_{ACB}\right)+ \frac{1}{2} \int d^3x\
\rho\left(\xi^i_{\{A}\xi^j_B\delta^{(1)}\Phi_{C\};ij} 
+ \xi^i_A\xi^j_B\xi^k_C\Phi_{;ijk} \right) \label{Kat1}
\end{eqnarray}
\end{widetext}
where $\Xi_{AB} = \xi^i_{A;k}\xi^k_{B;i}$, $\chi_{ABC}= \xi^i_{A;j}\xi^j_{B;k}\xi^k_{C;i}$ and $\Gamma_1 = \left. \partial \ln p/\partial \ln \rho\right|_s $ is the generalized adiabatic index governing the perturbations.
The notation $V_{\{ABC\}}$ used for example in the term \mbox{$\Xi_{\{AB}\nabla\cdot\bj{\xi}_{C\}}$} indicates that all cyclic permutations of the indices labeling the modes must be added, in other words  $V_{\{ABC\}}= V_{ABC}+V_{BCA}+ V_{CAB} $ .
 
Note that we have changed the sign of $\Phi$ from that used by Schenk et al \cite{Katrin1}, to be consistent with convention of Ipser and Lindblom \cite{Ipser} and Chandrasekhar
\cite{Ellips}.

We have chosen to investigate a particularly simple version
of the nonlinear coupling, namely incompressible {\it perturbations}
of a {\it uniformly dense} star. In principle, we could allow 
finite $\Gamma_1$ for the perturbations, but we have made a tacit
assumption that the perturbations and the star obey the same
zero entropy equation of state with a single, large $\Gamma_1$.
This is a convenient assumption for keeping our investigations
of the nonlinear network relatively simple.

\subsection{Nonlinear coupling for the incompressible star}
As
the adiabatic index $\Gamma_1 \rightarrow \infty$, the divergence of the Lagrangian displacement approaches zero ($\nabla \bj{\cdot\xi}
\rightarrow 0$). In the coupling coefficients, equation (\ref{Kat1}), these two terms appear as a product which has to be handled with care.  

To find an expression for $p (\Gamma_1-1)  \nabla
\bj{\cdot\xi} \approx p \Gamma_1  \nabla
\bj{\cdot\xi} $ note that to first order the Lagrangian perturbations of density ($\Delta \rho$) and pressure ($\Delta p$) obey  
$\Delta \rho/\rho = - \nabla \bj{\cdot\xi}$ and $\Delta p/p=
\Gamma_1\Delta \rho/\rho$ (equations~{(I14) and (I17)}  of \cite{Katrin1})
yielding 
\begin{equation}
p\Gamma_1  \nabla \bj{\cdot\xi}=  - \Delta p
\label{dP2}
\end{equation}
Note that the term proportional to $ \nabla\cdot\bj{\xi}_A \nabla\cdot\bj{\xi}_B \nabla\cdot\bj{\xi}_C$
vanishes because it is multiplied by  only two powers of $\Gamma_1$. Thus the incompressible coupling coefficient reduces to 
\begin{align}
\kappa_{ABC} =& -\frac{1}{2} \int d^3x\ \Xi_{\{AB}\Delta p_{C\}}
-  p\left(\chi_{ABC}+\chi_{ACB}\right)
\nonumber\\&
+  \frac{1}{2} \int d^3x\
\rho\left(\xi^i_{\{A}\xi^j_B\delta^{(1)}\Phi_{C\};ij}
+ \xi^i_A\xi^j_B\xi^k_C\Phi_{;ijk} \right)~. \label{Kat2}
\end{align}

\subsection{Coupling of Bryan modes}

The Bryan solution of the linear modes is presented in terms of two scalar potentials $\delta\Phi$ equation \eqref{eq:dPhi} and $\delta U$, equation \eqref{eq:dU}, from which the Lagrangian displacement $\bj{\xi}$ is constructed in equation   \eqref{eq:BryanEig}. It is in terms of these three quantities that we now  construct the incompressible coupling coefficients \eqref{Kat2}.

Substituting the expression for the Lagrangian pressure perturbation into equation \eqref{Kat2}, the coupling coefficient becomes

\begin{align}
\kappa_{ABC} =& -\frac{1}{2} \int d^3x\ \rho\Xi_{\{AB}\delta U_{C\}}
\tag{$\kappa_1$} 
\\
& -\frac{1}{2} \int d^3x\ \rho\Xi_{\{AB}\delta \Phi_{C\}}  
\tag{$\kappa_2$} 
\\
&+  \frac{1}{2} \int d^3x\
\rho\xi^i_{\{A}\xi^j_B\delta \Phi_{C\};ij}
\tag{$\kappa_3$} 
\\
& -\frac{1}{2} \int d^3x\
(\Xi_{\{AB}\bj{\xi}_{C\}})\cdot\nabla p-
p\left(\chi_{ABC}+\chi_{ACB}\right)
\tag{$\kappa_4$} 
\\
&+  \frac{1}{2} \int d^3x\
\rho\left( \xi^i_A\xi^j_B\xi^k_C\Phi_{;ijk} \right) 
\tag{ $\kappa_5$}\\
\label{CouplingCoeffs}
\end{align}
The domain of integration  for $\kappa_1,\ \kappa_2,\ $ and $\kappa_4$ is the volume of the star,  the Maclaurin spheroid of
eccentricity $e$.  The integrals for 
$\kappa_3$ and $\kappa_5$ are taken over all space; however the factor $\rho$  contains a step
function that limits the integrals to the volume of the star and possibly a
surface term. As written $\kappa_{ABC}$ has units of energy; computed values are appropriately non-dimensionalized by dividing by the mode energy, $\epsilon$.

These terms can be rewritten in various ways that facilitate, and check,
the calculation of the coupling coefficients. The only assumptions that are made
during these manipulations are that the pressure $p$ vanishes on the stellar surface, equation \eqref{eq:p},
and that the Lagrangian displacement has no divergence, $\nabla \cdot \bj{\xi}=0$.

Schenk et al \cite{Katrin1} also give an alternative expression for the coupling coefficients in equation (4.27) of their paper. For completeness the correspondence between the coupling terms $\kappa_1-\kappa_5$ and their expression is summarized in Appendix~\ref{sec:Coupcorresp}.

\subsubsection{ $\kappa_2$ and $\kappa_3$ }
\label{subkappa2_3}

The term $\kappa_3$ contains the second derivative of $\delta \Phi$. From equation \eqref{eq:dPhi}, this contains a  delta function and yields a surface integral term of the same form as $\kappa_5$.

Integrating \eqref{eq:dPhi} once 
 gives rise to the boundary condition (equation (3.11) in \cite{Ipser})
\begin{align}\left[n^i \nabla_i \delta\Phi \right]_{\outStar}- \left[n^i\nabla_i
\delta\Phi\right]_{\inStar} &=
-\left[ \frac{ 4 \pi G \rho^2}{ |\nabla p|}(\delta U+
\delta \Phi) \right]_{ \inStar} \notag\\
&=-\left[ 4 \pi G\rho \ \bj{n} \cdot   \bj{\xi} \right]_{ \inStar} \label{BoundaryII} 
\end{align}
where use was made of the $\Delta p = 0$ boundary condition to obtain the second equation, $\Sigma$ is a function that has a gradient equal to the outward directed unit normal vector $n_a$ to the star and vanishes  on the stellar surface. In particular 
\begin{equation}\Sigma = -\frac{p }{|\nabla p|} \end{equation} is an example of one such function.

The surface term in $\kappa_3$ can be computed by letting 
\begin{equation}\delta \Phi_{\ ;i} = H(-\Sigma)\delta
\Phi_{\inStar \ ;i}+H(\Sigma)\delta \Phi_{\outStar\ ;i}
\label{eq:Surk}
\end{equation}
 where $H(x)$ is the Heaviside unit step function.  Differentiating equation \eqref{eq:Surk} and using equation \eqref{BoundaryII} gives
\begin{align}\delta \Phi_{;ij} =
&H(-\Sigma)\delta
\Phi_{\inStar\ ;ij}+H(\Sigma)\delta \Phi_{\outStar\ ;ij}\notag\\
&-4 \pi
G\rho \delta(\Sigma) n_jn_in_k\xi^k  \label{dPhiii} \end{align} 

Using equation \eqref{dPhiii}, and integrating by parts once  $\kappa_2$ and  $\kappa_3$ can be rewritten as an integral over the stellar surface only: 
\begin{eqnarray}\kappa 2+\kappa 3
&=& -\frac{1}{2} \int  d^2 A_i\ \rho\left( \xi^i_{\{A;j}\xi^j_{B}\delta \Phi_{C\}\ } -\xi^i_{\{A}\xi^j_{B}\delta \Phi_{C\ ;j\}} \right) \nonumber\\
&&-\frac{3}{2} \int   d^2 A_i\  4 \pi G \rho^2  n_jn_k \xi^i_{A}\xi^j_{B}\xi^k_{C} 
\label{ktwoandthree} 
\end{eqnarray}
As we show in Appendix~\ref{sec:kappa5}  , the last term in equation \eqref{ktwoandthree} is equal to $(-3)\times \kappa_5$

\subsubsection{$\kappa_4$}

Term $\kappa_4$ is difficult to compute numerically as written. The two terms cancel to leading order in $\Omega^2$ in the spherical limit, $(e\rightarrow 0)$.  
Reorganizing the differentiation and using the  $\Delta p = 0$ boundary condition  allows us to rewrite $\kappa_4$ as a complete
divergence. The resulting  surface integral is the same order in $\Omega$ as the other terms in the coupling coefficient and can be integrated to yield a finite result.  The details of the calculation are given  in Appendix~\ref{AppendixKappa4} with the result
\begin{align}
\kappa_4 &= \frac{1}{2} \int d^2A_k \ \rho\left[(\delta U_B+\delta \Phi_B)\xi^k_{C;i} +(\delta U_C+\delta \Phi_C)\xi^k_{B;i}\right] \xi^i_A\notag\\
 & -\frac{1}{2} \int d^2A_k \ \frac{\rho}{|\nabla p|^2}(\delta U_A+\delta \Phi_A)p^{;k}\xi^i_B\xi^j_{C}p_{;i;j}\label{kappa4}
\end{align}
The result is asymmetric in the indices ${ABC}$, with no cyclic permutation of indices implied. (The expression can be symmetrized over the indices if desired by adding the two terms where the indices are cyclic permutations of ABC and dividing the complete expression by three.)

\subsubsection{ $\kappa_5$ }
\label{subseckappa5}
The term $\kappa_5$ scales as $O(\Omega^6)$ compared to the other terms that scale as $O(\Omega^2)$. Appendix~\ref{sec:kappa5} shows that $\kappa_5$ can be rewritten as the surface term.
\begin{equation}\kappa_5 = \frac{1}{2}\int d^3x \ 4\pi G\rho^2  \delta(\Sigma)\left(\xi_A^i\xi_B^j\xi_C^k n_kn_in_j\right)
\label{eq:k5} 
\end{equation}
In the spherical limit equation \eqref{eq:k5} is identically zero.

In the nonspherical case, this is the only term in the coupling coefficients that cannot be calculated by means of our polynomial representation and term-by-term integration (Section~\ref{CompMethod}). It does, however, yield an integral that can be solved analytically as is shown in Appendix~\ref{sec:kappa5}.

\section{Computation of Coupling Coefficients and other integrals}
\label{CompMethod}
The computation of the coupling coefficients, shear damping, gravitational radiation driving or damping and normalization require the integration of quantities related to $\delta U$, $\delta \Phi$, \bj{\xi} and their covariant derivatives to be computed over the volume of the background spheroid and its surface.  The highly oscillatory nature of the functions for large $n$ values precludes a direct numerical integration.

Our approach is to represent the functions analytically as far as possible.
We store the linear potentials, $\delta U$, $\delta \Phi$ as polynomials in $r$ and $\varpi$. These potentials can be constructed directly from the recursion relations in   $z$ and $\varpi$ as derived in Appendix~\ref{BryanRecursion},  if the frequency of the eigenmode is known. 
Integration of the coupling coefficient over the volume of the spheroid can then be achieved analytically.  The coupling coefficient is computed as a polynomial and each term integrated individually.

The analytic integral of a typical term in the coupling coefficient is obtained  by working in the coordinates $s$ and $\tilde{\theta}$ which map the interior of the star onto the unit sphere and are defined by the coordinate transformation.
\begin{equation}
\varpi = R_es \sin \tilde{\theta} \ \ z=R_ps \cos \tilde{\theta}
\end{equation}
In these coordinates the volume element is \mbox{ $d^3x = R_e^2R_ps^2 \sin \tilde{\theta}\ ds\ d\tilde{\theta}\ d\phi $} and the integral over the star can be written as  
\begin{align}
\int d^3x\ &  \varpi^kz^he^{i(m_A+m_B+m_C)\phi}\notag \\=&R_e^{2+k}R_p^{1+h}\int_0^{2\pi}e^{i(m_A+m_B+m_C)\phi }  d\phi \notag\\
&\times\int_0^\pi \sin^{k+1}\tilde{\theta} \ \cos^h\tilde{\theta} \  d\theta \int_0^1   s^{2+k+h} ds \notag\\
=& \frac{2\pi R_e^{2+k}R_p^{1+h} }{3+k+h} \delta (m_A+m_B+m_C)\notag\\
 &\times B\left(\frac{k+2}{2},\frac{h+1}{2}\right)\ (1+(-1)^h) 
\end{align}  
where $B(x,y)$ is the Beta function.

All other integrals over the volume of the star, for example the normalization, damping etc.  can be done in a similar way. 
The analytic result for normalization is used to check the accuracy of the method.

In spite of using the analytic value for the individual terms in the integral, round-off errors become a serious problem. For the case where the coefficient is stored as a double precision floating point number, all significant digits in the normalization calculation have been lost by the time $n>16$. Successive addition and subtraction of nearly equal terms cause the loss of precision. 

To overcome the loss in precision we made use of the GNU Multiple Precision Arithmetic Library GMP 4.1.2
% found at http://www.swox.com/gmp/ 
and worked to 32 digits accuracy.  This resulted in the normalization integral differing from the analytic value by $<10^{-25}$ and coupling coefficients obeying their selection rules listed in Table~\ref{tab:Selection} to within $10^{-19}$ . The cost of this increased accuracy is speed: $10^4$ couplings take approximately 4 days to compute on a Pentium 4, 2GHz processor.

\section{Properties of Coupling Coefficients}
\label{Selectionrules}

The importance of a coupling in an oscillator network is determined by two factors: the magnitude of the coupling $|\kappa_{\overline{\alpha}\beta\gamma}|$ and the detuning 
\begin{equation} \delta w_{\overline{\alpha}\beta\gamma}=w_\alpha -w_\beta - w_\gamma \end{equation}
or frequency mismatch between the three coupled modes with frequencies $w_\alpha$,  $w_\beta$ and $ w_\gamma$.  
Modes with small detuning will ultimately play a far more important role in the dynamics of the problem than those with a larger frequency mismatch. Basically,
a triplet with large detuning adds a very rapidly oscillating component to the nonlinear terms which on average will be zero. A small detuning results in a slowly varing contribution which may have a large accumulative effect and substantial influence.    

\begin{figure}[h]
\includegraphics[width=\columnwidth]{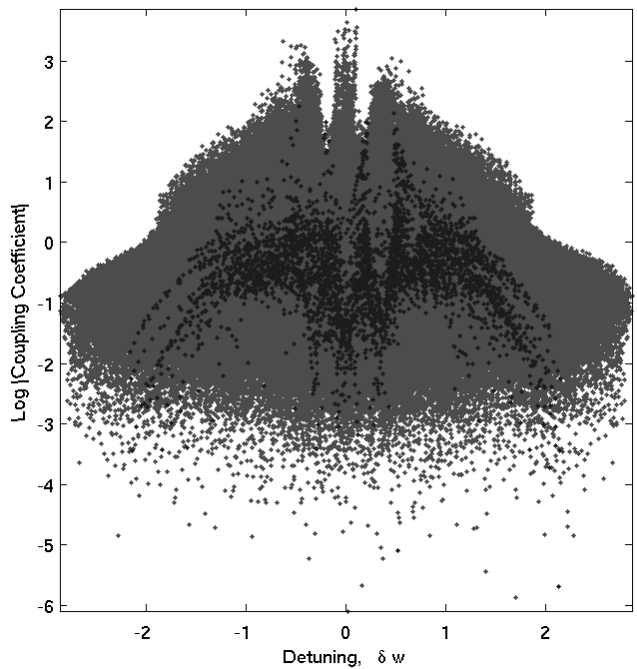}
\caption{(Color online) The magnitude of the coupling coefficient of a triplet of modes vs its detuning. In the background (red/light) are all the 766884 non-zero couplings for modes with $n\leq 12$.  The foreground (blue/dark) shows the 3748 direct couplings to the ($n=3$, $m=2$) r-mode.} 
\label{DogPlot}
\end{figure}

The computed values for all the coupling coefficients up to a maximum value of $n=12$ as a function of detuning are shown in Figure~\ref{DogPlot}.

As mentioned in Schenk et al~\cite{Katrin1}, the coupling coefficients obey a set of selection rules.  Our direct calculation verifies  those stated in equations (5.11), (5.12) and (5.13) of their paper.  An additional selection rule, the triangular selection rule on $n$, was discovered empirically.  A complete list of selection rules obeyed by the coupling coefficients in the spherical limit is summarized in Table~\ref{tab:Selection}. All other coupling coefficients are found to be nonzero.  We find that these selection rules persist for modes with nonzero eccentricity, at least for the  few coefficients we checked.

\begin{table}[h]
\begin{tabular}{||c|cr||}\hline\hline
Selection Rule & Statement of rule &\\ \hline
parity &$ n_A + n_B +n_C = \mbox{odd number}$ & $\Rightarrow \kappa_{ABC}=0$\\\hline
azimuthal, $m$ &$ m_A + m_B +m_C \neq 0$ & $\Rightarrow \kappa_{ABC}=0$\\\hline
triangular & $n_C\leq|n_A-n_B|$  &  $\Rightarrow \kappa_{ABC}=0$  \\
&  $n_A+n_B \leq n_C $ &  $\Rightarrow \kappa_{ABC}=0$  \\
\hline\hline
\end{tabular}
\caption{Selection rules obeyed by coupling coefficients}
\label{tab:Selection}
\end{table}

Schenk et al \cite{Katrin1} also list a number of selection rules to $O(\Omega)$. If we consider  $\bj{\xi}$  to be $O(1)$ then equation \eqref{eq:BryanEig} implies that $\delta U$ and
$\delta \Phi$ are $O(\Omega^2)$.  Since all the coupling terms except $\kappa 5$ are composed of terms such as  $\xi \xi \delta U$ or $\xi \xi \delta \Phi$, the lowest order coupling coefficient is $O(\Omega^2)$. Thus the additional selection rules in  \cite{Katrin1} are trivially satisfied since they are only valid to $O(\Omega)$.

The empirically discovered triangular selection rule on $n$ has also been seen in other calculations  dealing with 2D planetary Rossby waves \cite{Siberman}.  The coupling coefficients used in those calculations are less complicated than those considered in our case. 

In their paper on the saturation of the r-mode instability \cite{Saturation}, Arras et al. make a momentum conservation assumption in their equation (A2) to obtain the scalings for a cascade solution. We find that all the couplings on which they based their calculation are zero since the momentum conservation assumption corresponds to the equality case of the triangular selection rule. However immediately adjacent to those modes are modes which give non-zero couplings and for which their WKB solution may still be valid.

We find that the contribution to the coupling coefficients of terms containing perturbations to the gravitational potential are of the same order of magnitude as the hydrodynamic terms.  
This calls into question the applicability  of the Cowling approximation,  which is often assumed to apply, 
for modes where the time scales are long and gravity has plenty of time to act.

\section{Discussion}

This paper provides the necessary analytic background and the computational approach required for calculating the couplings between the inertial modes of an incompressible perfect fluid star. These couplings determine  the oscillator network of nonlinearly coupled modes that we shall use to explore the r-mode instability in neutron stars in Part II \cite{JdB2}. 

Although we have focused on the zero eccentricity limit in this paper,  the techniques described are not limited to the spherical case, and remain valid for arbitrary rotation. An investigation into the effect of rotation on the eigenfunctions and frequencies, and resulting couplings and damping of modes will be an interesting study in its own right.

A complete list of selection rules obeyed by the couplings is given in Table~\ref{tab:Selection}. These rules along with the frequency spectrum shown in Figure~\ref{Freqfig} determine the basic nature and connectivity of the oscillator network. This structure will influence the nature of the dynamics of energy transfer from the unstable large scale r-mode to the smaller damping scales and will set the maximum attainable amplitude of the r-mode. 
The numerical integration of this network, and a detailed look at the non-linear dynamics that result, will be discussed in detail in  a Paper II \cite{JdB2}. The simplest dynamical system resulting from this network, the three mode Hamiltonian system,  has already been used to analytically explain the dynamics of the catastropic decay of large amplitude r-modes observed in the hydrodynamical simulations \cite{JdB3}. It is expected that investigation of the larger network will further increase our understanding of the evolution of the unstable r-mode.  

Thanks to Harald Pfeiffer for suggestions on the coding of the classes and search algorithms used to manipulate and compute the coupling coefficients.

This research is supported in part by NSF grants AST-0307273, PHY-9900672 and PHY-0312072 at Cornell University.

\appendix

\section{Generating Eigenfunctions in cylindrical coordinates}
\label{BryanRecursion}

Our method of generating the potentials uses the recurrence relations of
Legendre functions, equations~\eqref{LegendreRecurr} to construct the potentials as polynomials in $r$ and $z$.

\begin{eqnarray}(n-m)P^m_n(x)&=& x(2n-1)P^m_{n-1}(x)-(n+m-1)P^m_{n-2}(x)\nonumber\\
P^m_m(x)&=&(-1)^m(2m-1)!!(1-x^2)^{\frac{m}{2}}\nonumber\\
P^m_{m+1}(x)&=&x(2m+1)P^m_m(x) \label{LegendreRecurr}
\end{eqnarray}
Using equations~\eqref{LegendreRecurr} we derive the recurrence relations for the products of Legendre functions needed in equations \eqref{dUEig} and \eqref{dPhiEig}.
Define
\begin{eqnarray}
I_n^m &=& P_n^m(x_1)P_n^m(x_2)\nonumber\\
I_m^m &=&\left[(2m-1)!!\right]^2\left[\sqrt{(1-x_1^2)(1-x_2^2)}\right]^m\nonumber\\
I_{m+1}^m&=&x_1x_2\left[2m+1\right]^2I_m^m
\end{eqnarray}
We shall also need the quantities
\begin{eqnarray}
G_n^m &=& x_1P_n^m(x_1)P_{n-1}^m(x_2)+x_2P_n^m(x_2)P^m_{n-1}(x_1)\nonumber\\
G_m^m&=&0\nonumber\\
G^m_{m+1}& =& (x_1^2+x_2^2)(2m+1)I_m^m
\end{eqnarray}
Using equations \eqref{LegendreRecurr}  we can
write
\begin{widetext}
\begin{align}
(n-m)^2I^m_n& =
x_1x_2(2n-1)^2I_{n-1}^m+(n+m-1)^2I_{n-2}^m-(2n-1)(n+m-1)G_{n-1}^m\notag\\
(n-m)G_n^m&=(x_1^2+x_2^2)(2n-1)I_{n-1}^m+\frac{(n+m-1)}{(n-m-1)}\left((n+m-2)G^m_{n-2}-2x_1x_2(2n-3)I_{n-2}^m\right)
\end{align}
\end{widetext}
Notice that the only combinations of coordinates that enter these relations for $\delta U$, equation~\eqref{dUEig}  with \mbox{($x_1 = \xi$, $x_2 = \tilde{\mu}$)} are
\begin{align}
\sqrt{(1-x_1^2)(1-x_2^2)}&=\frac{\varpi}{b}\nonumber\\
(x_1^2+x_2^2) &= 1-\frac{\varpi^2}{b^2}+\frac{z^2}{b^2d^2}\nonumber\\
 x_1x_2& =\frac{z}{bd}\ . 
\end{align}

The gravitational potential $\delta \Phi_{\inStar}$  inside the star, equation~\eqref{dPhiEig} ($x_1 = \mu$, $x_2 = i\zeta$) can be constructed using  
\begin{eqnarray}
\sqrt{(1-x_1^2)(1-x_2^2)}&=&\frac{\varpi}{a}\nonumber\\
(x_1^2+x_2^2) &=& 1-\frac{\varpi^2}{a^2}-\frac{z^2}{a^2}\nonumber\\
 x_1x_2& =&\frac{iz}{a} \ .
\label{dPhirecurr}
\end{eqnarray}

\section{Term $\kappa_4$ as a surface term}
\label{AppendixKappa4}

This appendix computes the surface integral version of $\kappa_4$ through successive integrations by parts and application of the boundary condition $\Delta p = 0$. The integrations by parts can be conducted by means of the manipulations listed in Table~\ref{tab:pressterms}.
\begin{table}[h]
\[
\renewcommand{\arraystretch}{1.2}
\begin{array}{||rcl||rcl||} \hline\hline
&&&V_0 &=& p \xi^i_{A;j}\xi^j_{B;k}\xi^k_{C;i}\\
V_0&=&D_1-V_1&
D_1&=& \left(p \xi^i_{A}\xi^j_{B;k}\xi^k_{C;i}\right)_{;j}\\
V_1& =&V_2+V_3& V_2&=&p_{;j}
\xi^i_{A}\xi^j_{B;k}\xi^k_{C;i}\\
&&& V_3& =&p \xi^i_{A}\xi^j_{B;k}\xi^k_{C;i;j}\\
V_2 &=& D_2-V_4&D_2 &=& \left(p_{;j}
\xi^i_{A}\xi^j_{B}\xi^k_{C;i}\right)_{;k}\\
V_4 &=& V_5+V_6& V_5&=& 
p_{;j;k}
\xi^i_{A}\xi^j_{B}\xi^k_{C;i}\\
&&&V_6&=&p_{;j}\xi^j_{B}\Xi_{AC}\\
V_3 &=&D_3-V_7&D_3& =& \left(p\xi^i_{A}\xi^j_{B;k}\xi^k_{C;j}\right)_{;i}\\
V_7&=&V_8+V_9&V_8&=&p_{;i}\xi^i_{A}\Xi_{BC}\\
&&&V_9& =& p\xi^i_{A}\xi^j_{B;k;i}\xi^k_{C;j}\\
V_5&=&D_4-V_{10}&D_4&=&\left(p_{;j;k}
\xi^i_{A}\xi^j_{B}\xi^k_{C}\right)_{;i}\\
V_{10}&=&V_{11}+V_{12}&V_{11}&=&p_{;j;k;i}
\xi^i_{A}\xi^j_{B}\xi^k_{C}  \\
&&&V_{12}&=&p_{;j;k}
\xi^i_{A}\xi^j_{B;i}\xi^k_{C}\\\hline\hline
\end{array}
\]
\caption{Terms generated when integrating $p\chi_{ABC}$ by parts. Terms labeled $D$ yield surface integrals while those labeled $V$ are volume integrals. The left hand column shows successive steps in the integration process while the right hand column summarizes the terms.
}\label{tab:pressterms}
\end{table}
Using Table~\ref{tab:pressterms} $\kappa_4$ can be rewritten as follows.  
\begin{align}
p\chi_{ABC}& = D_1-D_2 +V_5 +V_6-V_3 \notag\\
& = D_1-D_2 +D_4 +V_6-V_{11}-V_{12}-D_3+V_8+V_9 \ .
\end{align}
Let an over bar indicate that the B and C have been interchanged. Then
\begin{align}
p\chi_{ACB} & = \overline{D_1}-\overline{D_2} +\overline{V_5} +\overline{V_6}-\overline{V_{3}}\notag \\
& =\overline{D_1}-\overline{D_2}+V_{12}+\overline{V_6}-V_{9}  
\ .  \end{align}
For the case of an incompressible fluid the pressure is a quadratic function in $x$, $y$, and $z$ and as a result $V_{11} =0$. 
This allows $\kappa_4$ to be written as 
\begin{equation}\kappa_4=\frac{1}{2} \int d^3x\ \left( D_1-D_2+D_4-D_3+\overline{D_1}-\overline{D_2}\right)
\end{equation}
which is a complete divergence and can be expressed as a surface integral over the boundary of the star.  

The pressure vanishes on the stellar surface, so, the terms $D_1$,  $\overline{D_1}$ and $D_3$ can be ignored and
 \begin{equation}\kappa_4=\frac{1}{2} \int d^3x\ \left(-D_2+D_4-\overline{D_2}\right) \ .
\end{equation}
Finally the $\Delta p = 0$ boundary condition can be used to rewrite the nonzero terms:
\begin{widetext}
\begin{eqnarray}-\frac{1}{2} \int d^3x\ (D_2+\overline{D_2}) &=& \frac{1}{2} \int d^2A_k \ \rho\left[(\delta U_B+\delta \Phi_B)\xi^i_A\xi^k_{C;i} +(\delta U_C+\delta \Phi_C)\xi^i_A\xi^k_{B;i}\right]\\
\frac{1}{2} \int d^3x\ D_4 &=& -\frac{1}{2} \int d^2A_k \ \frac{\rho}{|\nabla p|^2}(\delta U_A+\delta \Phi_A)p^{;k}\xi^i_B\xi^j_{C}p_{;i;j}
\end{eqnarray}
\end{widetext}

\section{Analytic computation of $\kappa_5$}
\label{sec:kappa5}

The computation of $\kappa_5$ requires a more detailed look at the background gravitational potential of a Maclaurin spheroid. We provide this background information in subsection~\ref{EllipBack}.  The integral expression for $\kappa_5$ in terms of the Lagrangian displacement is derived in subsection~\ref{k5int} while the analytic evaluation of this integral in terms of Legendre functions is detailed in subsection~\ref{k5ana}.  

\subsection{Background - Ellipsoidal Figures of Equilibrium}
\label{EllipBack} 
The background potential of a uniformly rotating star has been extensively studied.  The results presented here are a brief summary of the parts of Chandrasekhar's book \cite{Ellips} relevant to the computation of $\kappa_5$

The gravitational potential obeying the equation \mbox{$\nabla^2 \Phi = -4 \pi G \rho$} at an internal point $x_i$ (cartesian coordinates) of a homogeneous ellipsoid
with semi-axis $a_i$ is given by (equations (18),(40) of \cite{Ellips})
\begin{equation} 
\Phi_{\inStar} = \pi G \rho \left(I-\sum_{i=1}^3 A_ix_i^2\right)
\end{equation}  
where 
\begin{equation}
A_i = a_1a_2a_3\int_0^{\infty}\frac{du}{\Delta(a_i^2+u)}\end{equation} 
with
$\Delta(u)^2=(a_1^2+u)(a_2^2+u)(a_3^2+u)$ and \mbox{$I = a_1a_2a_3\int_0^{\infty}du / \Delta$}.

The gravitational potential at an external point can be found using equations (70),(71),(72) of \cite{Ellips}
and is 

\begin{equation}
\Phi_{\outStar} = \pi G \rho a_1a_2a_3\int_\lambda^{\infty} \frac{du}{\Delta}\left(1-\sum_{i=1}^3 \frac{x_i^2}{a_i^2+u}
 \right)
\end{equation}
with $\lambda$  the ellipsoidal
coordinate that is the positive root of the equation 
\begin{equation} 
\sum_{i=1}^3 \frac{x_i^2}{a_i^2+\lambda}  =1
\label{eq:lambda}
\end{equation}

The gradients of the potentials are (no summation over repeated indices)
\begin{eqnarray}
\Phi_{\inStar\ ;i} &=& -2\pi G \rho x_iA_i\\
\Phi_{\outStar\ ;i}&=& -2\pi G \rho x_i a_1a_2a_3\int_\lambda^\infty
\frac{du}{\Delta(a_i^2+u)}
\end{eqnarray}

Note that on the surface of the star corresponding to $\lambda =0$  the inner and
outer potentials and their gradients are continuous.
   
\subsection{Integral Expression for Term $\kappa_5$}
\label{k5int}

The coupling coefficients require the term $\Phi_{;ijk}$
multiplied by the density of the star to be known.  The potential in the
interior of the star is a quadratic function of $x,y$ and $z$ so it will
vanish after taking three derivatives or $\Phi_{\inStar\ ;ijk}=0$ . The density outside the star also
vanishes.  The only remaining  contribution is due to a surface integral
originating from the discontinuity in $\Phi^{;ij}$ on the boundary which gives rise to a
$\delta$ function when taking the third derivative. To compute this term express the second derivative of the potential as 
\begin{equation}\Phi_{;ij} = H(-\Sigma)\Phi_{\inStar\ ;ij}+ H(\Sigma)\Phi_{\outStar\ ;ij} \ .
\end{equation}
Then
\begin{equation}
\begin{split}
\Phi_{;ijk} =  H(-\Sigma)\Phi_{\inStar\ ;ijk}+ H(\Sigma)\Phi_{\outStar\ ;ijk} \\+
n_k\delta(\Sigma)\left(\Phi_{\outStar\ ;ij} - \Phi_{\inStar\ ;ij}\right) 
\end{split}
\end{equation}
where the first term is zero and the second does not contribute anything to the
coupling coefficient.  What remains is to calculate $\Phi_{\outStar\ ;ij} - \Phi_{\inStar\ ;ij}$ on the boundary of the star:
\begin{align}
(\Phi_{\outStar\ ;ij} - \Phi_{\inStar\ ;ij})|_{\lambda = 0} &= 2\pi G \rho\frac{
a_1a_2a_3 x_i}{a_i^2\Delta(0)}\frac{\partial\lambda}{\partial x^j}\Biggr|_{\lambda=0}\notag\\
&= 2\pi G \rho \frac{x_i}{a_i^2}\frac{\partial\lambda}{\partial x^j}\Biggr|_{\lambda=0} \label{Phiijtemp}
\end{align}
Differentiating equation (\ref{eq:lambda}) with respect to $x^j$ and subsequently specializing to case of the stellar boundary gives 
\begin{eqnarray}
\frac{2x_j}{a_j^2+\lambda} &=&\frac{\partial\lambda}{\partial
x^j} \sum _{i=1}^{3}\frac{x_i^2}{(a_i^2+\lambda)^2}\nonumber\\
\frac{\partial\lambda}{\partial
x^j}\Biggr|_{\lambda=0}&=&\frac{2x_j}{a_j^2}\frac{1}{\sum _{i=1}^{3}x_i^2/a_i^4}
\label{dlambda}
\end{eqnarray} 
Substituting \eqref{dlambda} into \eqref{Phiijtemp} gives
\begin{align}
(\Phi_{\outStar\ ;ij} - \Phi_{\inStar\ ;ij})|_{\lambda = 0}&=\pi G \rho \frac{2
x_i}{a_i^2}\frac{2x_j}{a_j^2}\frac{1}{\sum _{i=1}^{3} x_i^2/a_i^4}\notag\\ & = 4
\pi G \rho n^i n^j
\end{align}
Thus $\kappa_5$ due to the surface term is
\begin{eqnarray}\kappa_5 = \frac{1}{2}\int d^3x\ 4\pi G\rho^2  \delta(\Sigma)\left(\xi_A^i\xi_B^j\xi_C^k n_kn_in_j\right)
\label{eq:k5int}
\end{eqnarray}

\subsection{Analytic intergration of $\kappa_5$}
\label{k5ana}
 The integral for $\kappa_5$ \eqref{eq:k5int} can be computed analytically by substituting the explicit expressions for \bj{\xi} and $\delta U$ and working in spheroidal coordinates on the surface of the star. After some manipulation \cite{JdBThesis} equation~\eqref{eq:k5int} reduces to: 
 \begin{equation}
\begin{split}
\kappa_5  
= \frac{2\pi G\rho^2 }{a^2 (4\Omega^2)^3  } \left[ \frac{\beta}{w\sqrt{(1-w^2)}} \left(\frac{(1-\xi_0^2)}{P^m_n(\xi_0) }\frac{d P^m_n}{d \xi_0}
-\frac{\xi_0 m}{w}\right)\right]_A \\  
 [\cdots]_B\ [\cdots]_C\ \ \delta(  m_A - m_B- m_C)   \\
\int_{-1}^1 d\mu \frac{1}{(\mu^2 +\zeta_0^2)} \ P^{m_A}_{n_A}(\mu) \ P^{m_B}_{n_B}(\mu)\ P^{m_C}_{n_C}(\mu)
\end{split}
\end{equation}
This expression is symmetrical with respect to the mode indices, $A$, $B$ and $C$ and the term $[\cdots]_B$ indicates that a similar expression as that contained in the square braces $[\cdots]_A$ for mode $A$ should be included.  Note that it is these three factors that determine the $O(\Omega ^6)$ scaling of $\kappa_5$, while multiplication by the normalization factors, $\psi$ of each mode, cancels the $ \Omega ^6$ in the denominator of the constant factor.

To compute the contribution of the integral, suppose $u=v+w$,  $\ 0<l,n,m$, and $|u|\leq l$,  $|v|\leq m$,  $|w|\leq n$. Then the integral contribution rewritten in terms of partial fractions is  
\begin{align}
\int_{-1}^1 d\mu \frac{P_l^uP_m^vP^w_n}{(\mu^2 +\zeta_0^2)}= \int_{-1}^{1}d\mu  \frac{P_l^uP_m^vP^w_n}{2i\zeta_0} \left(\frac{1}{\mu -i\zeta_0}-\frac{1}{\mu +i\zeta_0}\right)
\end{align}
The solution to the general case can be built up using induction. 
This method for calculating the integral is based on the approach used by Gaunt \cite{Gaunt} to calculate the integral of the product three Legendre functions.
We first consider the two base cases: $n=w$ and   $n=w+1$.

Case $n=w$:
Recall that 
\begin{align}
P^n_n  &=   \frac{(-1)^n(2n)!}{ 2^nn!} (1-x^2)^{n/2} \notag\\
P^{-m}_n &= (-1)^m\frac{(n-m)!}{(n+m)!} P^m_n 
\end{align}  (also holds for $Q^m_n$) and use the integral \cite{MacRobert}\cite{Glasgow} \begin{equation}\int_{-1}^{1} dx \ x^k\frac{(1-x^2)^{m/2}}{(z-x)} P^m_n = (-2i)^m(1-z^2)^{m/2}Q^m_n(z)z^k
\label{eq:McR}\end{equation}
 provided $m\leq n$; $k=0,1,\dots,n-m$; $z$ is in the complex plane with a cut along the interval $(-1,1)$ on the real axis. So
\begin{equation}
\begin{split}
\int_{-1}^1 d\mu \frac{P_l^uP_m^vP^n_n}{(\mu^2 +\zeta_0^2)}= \frac{(2n)!}{2^nn!(2i\zeta_0)}\int_{-1}^{1} \left(\frac{1}{\mu -i\zeta_0}-\frac{1}{\mu +i\zeta_0}\right)\\ \times \left[ (1-\mu^2)^{\frac{u}{2}}  P_l^{u} \right]\left[(1-\mu^2)^{-\frac{v}{2}} P_m^{v} \right] d\mu
\end{split}
\end{equation}
Using Rodriques's Formula 
\begin{align}
P_n(x) &= \frac{1}{2^nn!} \frac{d^n(x^2-1)^n}{dx^n} \notag\\ 
P^m_n(x) &= (-1)^m(1-x^2)^{\frac{m}{2}}\frac{d^mP_n}{dx^m}
\end{align}
it can be seen that  
the second term is a polynomial of at most degree  $m-v$.
So if $m-v \leq l-u$ the integral can be evaluated using equation (\ref{eq:McR}) to give 
\begin{widetext}
\begin{eqnarray}
\int_{-1}^1 d\mu \frac{P_l^uP_m^vP^n_n}{(\mu^2 +\zeta_0^2)}&=& -(-2i)^l Q^u_l(i\zeta_0)P^v_m(i\zeta_0)P^n_n(i\zeta_0)\left(1+(-1)^{l+m+n}\right) 
\label{eq:QPPnn}
\end{eqnarray}
Here use was made of $P^m_n(-z)=(-1)^nP^m_n(z)$ and $Q^m_n(-z)=-(-1)^nQ^m_n(z)$. 

Case $w=n-1$, $u=v+n-1$: 
Use $P^{n-1}_n = (2n-1) xP^{n-1}_{n-1}$, so that 
\begin{eqnarray}
\int_{-1}^1 d\mu \frac{P_l^uP_m^vP^{n-1}_n}{(\mu^2 +\zeta_0^2)}&=& \frac{1}{(2i\zeta_0)}\int_{-1}^{1} \left(\frac{1}{\mu -i\zeta_0}-\frac{1}{\mu +i\zeta_0}\right)P_l^{u}P_m^{v}(2n-1)\mu P^{n-1}_{n-1}\nonumber\\
&=& \frac{(2n-1)i\zeta_0}{(2i\zeta_0)}\int_{-1}^{1} \left(\frac{1}{\mu -i\zeta_0}+\frac{1}{\mu +i\zeta_0}\right)P_l^{u}P_m^{v} P^{n-1}_{n-1}\nonumber\\
&=&-(-2i)^l Q^u_l(i\zeta_0)P^v_m(i\zeta_0)(2n-1)(i\zeta_0)P^{n-1}_{n-1}(i\zeta_0)\left(1+(-1)^{l+m+n}\right)\nonumber\\
&=&-(-2i)^l Q^u_l(i\zeta_0)P^v_m(i\zeta_0)P^{n-1}_{n}(i\zeta_0)\left(1+(-1)^{l+m+n}\right)
\end{eqnarray} 
General case  $m-v \leq l-u$, by induction: Suppose for all $k$, $w< k < n$ 
\begin{eqnarray}\int_{-1}^1 d\mu \frac{P_l^uP_m^vP^w_k}{(\mu^2 +\zeta_0^2)}=-(-2i)^m Q^u_l(i\zeta_0)P^v_m(i\zeta_0)P^{w}_{k}(i\zeta_0)\left(1+(-1)^{l+m+k}\right)\end{eqnarray}
Then using the recursion relation, equation~\eqref{LegendreRecurr} (a similar expression holds for $Q^m_n$)
%   $P^w_n = \frac{2n-1}{n-w}xP^w_{n-1}-\frac{n+w-1}{n-w}P^w_{n-2}$ and $Q^w_n = \frac{2n-1}{n-w}xQ^w_{n-1}-\frac{n+w-1}{n-w}Q^w_{n-2}$ 
it can be shown that
\begin{eqnarray}
\int_{-1}^1 d\mu \frac{P_l^uP_m^vP^w_n}{(\mu^2 +\zeta_0^2)}
&=& \frac{1}{(2i\zeta_0)}\int_{-1}^{1} \left(\frac{1}{\mu -i\zeta_0}-\frac{1}{\mu +i\zeta_0}\right)P_l^{u}P_m^{v}\left(\frac{2n-1}{n-w}\mu P^w_{n-1}-\frac{n+w-1}{n-w}P^w_{n-2}\right)\nonumber\\
&=& \frac{
(2n-1)i\zeta_0}{(n-w)(2i\zeta_0)}\int_{-1}^{1} \left(\frac{1}{\mu -i\zeta_0}+\frac{1}{\mu +i\zeta_0}\right)P_l^{u}P_m^{v} P^{w}_{n-1} \nonumber\\
&&+ \left( (-2i)^l Q^u_l(i\zeta_0)P^v_m(i\zeta_0)P^{w}_{n-2}(i\zeta_0)\left(1+(-1)^{l+m+n-2}\right)\frac{n+w-1}{n-w}P^w_{n-2} \right)  \nonumber\\
&=&- (-2i)^l Q^u_l(i\zeta_0)P^v_m(i\zeta_0)\left(1+(-1)^{l+m+n}\right)
\left(\frac{
(2n-1)i\zeta_0}{(n-w)}P^{w}_{n-1}-\frac{n+w-1}{n-w}P^w_{n-2} \right)  \nonumber\\
&=&- (-2i)^l Q^u_l(i\zeta_0)P^v_m(i\zeta_0)P^w_n(i\zeta_0)\left(1+(-1)^{l+m+n}\right) \nonumber\\
\end{eqnarray}
completing the proof by induction since the base cases of $n=w$ and $n=w+1$ were proved first.
\end{widetext}
For the other cases $m-v>l-u$ etc, there is nothing in this derivation that requires $v$, $u$ to be positive. Thus you could just relabel the Legendre functions appropriately and do the same calculation. The $Q$ factor in the answer then always corresponds to the original $P$ with largest difference between order and degree.

\section{ (4.27) Coupling Coefficients of Schenk et al.}
\label{sec:Coupcorresp}
Here we clarify the correspondence between the $\kappa_1-\kappa_5$ of \eqref{CouplingCoeffs} and those in equation
(4.27) of \cite{Katrin1}, which is reproduced in equation \eqref{kap4_27} below. The terms containing $\di\bj{\xi}$ have been omitted.
The $\delta p$ term has also been replaced using equation~\eqref{Eulerian_p}.

\begin{align}
\kappa_{ABC} =& -  \frac{1}{2} \int d^3x\ \xi^i_{\{A}\xi^j_B\delta (\rho\delta
U_{C\}})_{;ij} \tag{ $\overline{\kappa_1}$} \\
&  -  \frac{1}{2} \int d^3x\xi^i_{\{A}\xi^j_B\delta (\rho\delta
\Phi_{C\}})_{;ij}\tag{ ${\overline\kappa_2}$} \\
&+  \frac{1}{2} \int d^3x\
\rho\xi^i_{\{A}\xi^j_B\delta \Phi_{C;ij\}} \tag{ $\overline{\kappa_3}$} \\
& -  \frac{1}{2} \int d^3x\
 \xi^i_A\xi^j_B\xi^k_C p_{;ijk}\tag{ $ \overline{\kappa_4}$} \\
&+  \frac{1}{2} \int d^3x\
\rho \xi^i_A\xi^j_B\xi^k_C\Phi_{;ijk} \tag{ $\overline{\kappa_5}$}\\
\label{kap4_27}
\end{align}

In this case the domain of integration is all space and the pressure is taken to
be $p=p H(-\Sigma)$, that is, it vanishes outside the star. All fields are taken to
vanish at infinity, and complete divergence terms can be ignored. So doing
integration by parts twice yields $\overline{\kappa_1} = \kappa_1$ and
$\overline{\kappa_2}=\kappa_2$, the step function in $\rho$ restricting the
integration to the volume of the star only. 
Note that $\overline{\kappa_3} = \kappa_3$ and
$\overline{\kappa_5}=\kappa_5$ trivially.

The $\overline{\kappa_4}$ term can be treated using the machinery developed in
Table~\ref{tab:pressterms}:  
\begin{align}
\xi^i_A\xi^j_B\xi^k_Cp_{;ijk} =& V_{11} \notag \\
=& -p\chi_{ABC} + D_1-D_2 +D_4 -D_3 \notag \\&+V_6-V_{12}+V_8+V_9  
\end{align}
and 
\begin{equation}
V_{9}-V_{12} =-p\chi_{ACB}+\overline{D_1}-\overline{D_2}+\overline{V_6}  
\end{equation}
so
\begin{equation}
\begin{split} \xi^i_A\xi^j_B\xi^k_Cp_{;ijk}=-p(\chi_{ABC}+\chi_{ACB}) + D_1-D_2 +D_4  \\ +\overline{D_1}-\overline{D_2}-D_3
+V_6+V_8+\overline{V_6}\end{split}
\end{equation}

All $D$ terms can be ignored because the boundary is taken to be at infinity. Using the definitions of Table~\ref{tab:pressterms}:
\begin{equation}V_6+V_8+\overline{V_6}=
\Xi_{\{AB}\xi_{C\}}^ip_{;i}.\end{equation} 

The only term that needs interpreting is $p_{;j}$  
\begin{equation}p_{;j} =
(H(-\Sigma)p)_{;j}=H(-\Sigma)p_{;j}-\delta(\Sigma)p n_j\end{equation} but since $p$ vanishes on
the boundary, we are left with
$H(-\Sigma)p_{;j}$ which restricts the integral to the volume of the star and yields an exact equivalence to the $\kappa_4$ term.

\bibliographystyle{apsrev}
\bibliography{../Paper1}

\end{document}